\documentclass{llncs}
\usepackage{makeidx}
\usepackage{amsfonts}
\usepackage{amsmath}
\usepackage{amssymb}
\usepackage[dvips]{graphicx}
\usepackage{algorithmic}
\usepackage{algorithm}
\usepackage{compress}

\makeatletter
\def\thm@space@setup{%
  \thm@preskip=\smallskip \thm@postskip=\smallskip
}
\makeatother

\spnewtheorem{observation}[problem]{Observation}{\bfseries}{\itshape}
\newcommand{\heading}[1]{\smallskip\par\noindent{\bf #1}}

\newenvironment{packed_enum}{
	\begin{enumerate}
		\setlength{\itemsep}{1pt}
	    \setlength{\parskip}{0pt}
		\setlength{\parsep}{0pt}
}{\end{enumerate}}

\newenvironment{packed_itemize}{
	\begin{itemize}
		\setlength{\itemsep}{1pt}
	    \setlength{\parskip}{0pt}
		\setlength{\parsep}{0pt}
}{\end{itemize}}

\def\O{\mathcal{O}{}}
\def\blt{\blacktriangleleft}

\def\wlt{\vartriangleleft}

\def\wle{\trianglelefteq}

\def\LH{{\rm LH}}
\def\UH{{\rm UH}}
\def\rset{{\mathfrak{Rep}}}

\def\computationproblem#1#2#3{
	\begin{center}
	\begin{tabular}{rp{9cm}}
	{\bf Problem:\enspace}&#1\\
	{\bf Input:\enspace}&#2\\
	{\bf Output:\enspace}&#3\\
	\end{tabular}
	\end{center}
}

\def\medcap{\mathord{\raisebox{0.02em}{\scalebox{1.1}{\ensuremath{\cap}}}}}
\def\medcup{\mathord{\scalebox{1.1}{\ensuremath{\cup}}}}

\def\recog{\textsc{Recog}}
\def\ext{\textsc{RepExt}}
\def\brep{\textsc{BoundRep}}
\def\sim{\textsc{Sim}}
\def\inclusion{\textsc{Inclusion}}
\def\sub{\textsc{SubSet}}
\def\super{\textsc{SuperSet}}
\def\int{\hbox{\bf \rm \sffamily INT}}
\def\pint{\hbox{\bf \rm \sffamily PROPER INT}}
\def\uint{\hbox{\bf \rm \sffamily UNIT INT}}

\def\cP{\hbox{\rm \sffamily P}}
\def\cNP{\hbox{\rm \sffamily NP}}

\def\cFPT{\hbox{\rm \sffamily FPT}}

 \def\calB{{\cal B}} \def\calC{{\cal C}}

   \def\calP{{\cal P}}
 \def\calR{{\cal R}}

\def\cp{{\rm cp}}
\def\L{\mathfrak{L}}
\def\R{\mathfrak{R}}

\begin{document}

\mainmatter

\title{Bounded Representations of Interval\\and Proper Interval Graphs\thanks{%
The first two authors are supported by ESF Eurogiga project GraDR as GA\v{C}R GIG/11/E023.}}
\author{Martin Balko\thanks{%
			Department of Applied Mathematics, Faculty of Mathematics and
	   		Physics, Charles University, Malostransk{\'e} n{\'a}m{\v e}st{\'\i} 25,
            118 00 Prague, Czech Republic. E-mail: \texttt{balko@kam.mff.cuni.cz}.}
		\and Pavel Klav\'{\i}k\thanks{%
			Supported by Charles University	as GAUK 196213.
			Computer Science Institute, Faculty of Mathematics and
	   		Physics, Charles University, Malostransk{\'e} n{\'a}m{\v e}st{\'\i} 25,
            118 00 Prague, Czech Republic. E-mail: \texttt{klavik@iuuk.mff.cuni.cz}.}
		\and Yota Otachi\thanks{School of Information Science, Japan Advanced Institute of
		Science and Technology. Asahidai 1-1, Nomi, Ishikawa 923-1292, Japan. Email:
		\texttt{otachi@jaist.ac.jp}}
		}
\titlerunning{Bounded Representations of Interval Graphs}
\authorrunning{M.~Balko, P.~Klav\'\i k, Y.~Otachi}
\institute{}

\maketitle

\begin{abstract}
Klav\'{\i}k et al.~[arXiv:1207.6960] recently introduced a generalization of recognition called the
\emph{bounded representation problem} which we study for the classes of interval and proper interval
graphs. The input gives a graph $G$ and in addition for each vertex $v$ two intervals $\L_v$ and
$\R_v$ called \emph{bounds}. We ask whether there exists a bounded representation in which each
interval $I_v$ has its left endpoint in $\L_v$ and its right endpoint in $\R_v$. We show that the
problem can be solved in linear time for interval graphs and in quadratic time for proper interval
graphs.

\hskip 2em Robert's Theorem states that the classes of proper interval graphs and unit interval graphs are
equal. Surprisingly the bounded representation problem is polynomially solvable for proper interval
graphs and \cNP-complete for unit interval graphs [Klav\'{\i}k et al., arxiv:1207.6960]. So unless
$\cP = \cNP$, the proper and unit interval representations behave very differently.

\hskip 2em The bounded representation problem belongs to a wider class of restricted representation
problems. These problems are generalizations of the well-understood recognition problem, and they
ask whether there exists a representation of $G$ satisfying some additional constraints. The bounded
representation problems generalize many of these problems.
\end{abstract}

\section{Introduction} \label{sec:intro}

In the recent data-filled world, visualization and graph drawing is becoming an increasingly more
important topic. One is frequently asked to work with a huge object and to understand its structure.
In some cases, it is useful to visualize the object in a way which reveals its structure.  A prime
example of this is the class of \emph{interval graphs} which is one of the oldest and
best-understood graph classes. An interval graph $G$ can contain many edges, so a standard drawing
is not very understandable. But it has an \emph{interval representation} $\calR$ which is a
collection of closed intervals $\{I_v : v \in V(G)\}$ representing the vertices of the graph such
that $I_u \cap I_v \ne \emptyset$ if and only if $uv \in E(G)$. This representation nicely describes
the structure of the edges. We denote the class of interval graphs by \int.

Interval graphs were first introduced by Haj\'os~\cite{hajos_interval_graphs} in 1957. They caught
quickly an attention of many researchers, for instance Benzer~\cite{benzer_interval_graphs} used them
in his experimental study of the DNA structure. The first polynomial-time recognition algorithms were
given already in 1960's~\cite{gilmore64,maximal_cliques}. After a decade, Booth and
Lueker~\cite{PQ_trees} finally described a linear-time recognition algorithm based on a new
tree-structure called PQ-trees, applicable also to other problems such as planarity. Nowadays, there
are over several hundred papers dealing with many aspects of interval graphs. 

An interval representation is called \emph{proper} if $I_u \subseteq I_v$ implies $I_u = I_v$, i.e.,
no interval is a proper subset of another interval. And it is called \emph{unit} if all intervals
are of unit length. We consider two important subclasses of interval graphs: \emph{proper
interval graphs} (\pint) are graphs which admit proper interval representations, and similarly for
\emph{unit interval graphs} (\uint). The well-known theorem of Roberts~\cite{roberts_theorem} states
that $\pint = \uint$.

\subsection{The Bounded Representation Problem}

Several recent papers study \emph{restricted representation problems} in which we ask whether there
exists, say, an interval representation of an input graph $G$ satisfying some additional
constraints; see for
example~\cite{kkv,simultaneous_interval_graphs,kkkw,blas_rutter,int_lengths_intersections}.  In this
paper, we study for the classes \int\ and \pint\ one such problem called \emph{bounded
representation}, recently introduced by Klav\'{\i}k et al.~\cite{kkorssv}. This problem is related
to many other restricted representation problems; see Section~\ref{sec:other_problems} for details.

For an arbitrary interval $I$, we denote its left endpoint by $\ell(I)$ and its right endpoint by
$r(I)$. Let $\L_v$ and $\R_v$ be two intervals defined for each $v \in V(G)$. A representation $\calR$ is
called a \emph{bounded representation} if $\ell(I_v) \in \L_v$ and $r(I_v) \in \R_v$ for each $v \in
V(G)$. The bounded representation problem is the following decision problem:

\computationproblem
{The Bounded Representation Problem -- $\brep(\calC)$}
{A graph $G$ and two intervals $\L_v$ and $\R_v$ for each $v \in V(G)$.}
{Is there a bounded representation $\calR$ of the class $\calC$?}

\begin{figure}[t!]
\centering
\includegraphics[scale=0.82]{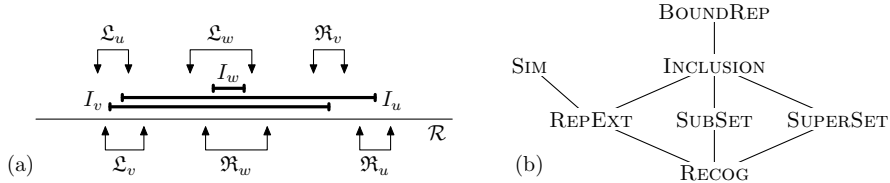}
\caption{(a) A bounded representation $\calR$ of the class \int\ is given for a graph $K_3$. 
There exists no bounded proper interval representation since $I_w$ is
always a proper subset of $I_u$ and $I_v$. (b) The Hasse diagram for different restricted
representation problemsi. If $\calP \le \calP'$, then the problem $\calP$ can be solved using the
problem $\calP'$.}
\label{fig:common}
\end{figure}

\noindent In the further text, we refer to the intervals $\L_v$ as \emph{left bounds} and to the
intervals $\R_v$ as \emph{right bounds}, or just simply \emph{bounds}. See
Fig.~\ref{fig:common}a for an example of an \brep\ instance. It is easy to see that the bounded
representation problem generalizes recognition; if all the bounds are set to $(-\infty,+\infty)$,
they pose no restriction at all. We also allow trivial bounds consisting of single points.

\subsection{Other Restricted Representation Problems} \label{sec:other_problems}

We review other restricted representation problems and discuss their relation to the bounded
representation problem. The problems were considered for different intersection classes of graphs
which we do not define formally; see the references for details. We note that all these problems
generalize the recognition problem (\recog). See Fig.~\ref{fig:common}b for the relations
between the problems.

\heading{Partial Representation Extension.} This problem denoted by \ext\ was introduced by
Klav\'{\i}k et al.~\cite{kkv}. The input prescribes together with $G$ an intersection representation
$\calR'$ of an induced subgraph $G'$. The goal is to find a representation $\calR$ of the entire $G$
which extends $\calR'$, i.e., it assigns the same sets to the vertices of $G'$ as $\calR'$. The
problem can be solved in polynomial time for interval graphs~\cite{kkv,blas_rutter,kkosv}, proper
and unit interval graphs~\cite{kkorssv}, function and permutation graphs~\cite{kkkw} and circle
graphs~\cite{cfk}. For chordal graphs in the setting of subtree-in-a-tree graphs, several versions
of the problem were considered in~\cite{kkos}, and almost all of them are \cNP-complete. It is known
that planar graphs have several intersection representations (contact representations of discs,
etc.), but extending these representations is \cNP-hard~\cite{int_planar_hard}.

The bounded representation problem generalizes partial representation extension, since one can
prescribed singleton bounds for the intervals of $G'$ and $(-\infty,+\infty)$ for the remaining
bounds. (We note that the bounded representation problem can be considered also for many other
classes of graphs.)

\heading{Inclusion Restrictions.} Function graphs are intersection graphs of continuous functions
defined on $[0,1]$. In \cite{kkkw}, the following problem was considered and proved to be \cNP-complete.
The input prescribes some functions partially, i.e., on partial domains $[a,b] \subseteq [0,1]$.
The goal is to extend them to the full domain $[0,1]$.

We consider more generally three different problems \inclusion, \sub, and \super\ for interval
graphs. In all problems, the input gives two intervals $A_v$ and $B_v$ for each vertex $v \in V(G)$.
The goal is to construct a representation such that $A_v \subseteq I_v \subseteq B_v$. Further for
\sub, we put all $A_v = \emptyset$, and for \super, we put all $B_v = (-\infty,\infty)$. It is easy
to see that these problems can be reduced to the bounded representation problems, and \inclusion\
can solve \ext.

\heading{Simultaneous Representations.} This problem denoted by \sim\ was introduced and solved for
several classes by Jampani et al.~\cite{simultaneous_interval_graphs}. The input consists of two
graphs $G_1$ and $G_2$ with some common vertices. The goal is to construct their representations
$\calR_1$ and $\calR_2$ such that the common vertices are represented the same. Bl\"asius et
al.~\cite{blas_rutter} reduce $\ext(\int)$ to $\sim(\int)$, and thus solve the first problem in
linear time. On the other hand, when the problem is generalized to $k$ input graphs, the best known
result for many classes is an \cFPT\ algorithm in the number of common vertices based on the partial
representation extension~\cite{kkv,cfk}. We are not aware of any relation of the simultaneous
representations problem to the other considered problems.


\heading{Motivation.} There are two very good motivations for studying the restricted representation
problems. The first motivation is that they might be applicable. For instance, one might want to
construct some specific representation of the given graph $G$. Using these restrictions, one can
force the representation to be constructed in this way. The other motivation is that to solve these
problems much better structural understanding is required. For classes like interval graphs, the
structure of all representations is well understood and one can just use PQ-trees to solve the
problems. For other classes like unit interval graphs~\cite{kkorssv} or circle graphs~\cite{cfk}, the
new structural results were developed which might be fruitful also for other purposes. In
mathematics, it is generally desirable to have problems which force one to get better understanding
of the objects.

\subsection{Our Results}

In this paper, we prove the following two theorems. For $\brep(\int)$, we assume that the endpoints
of the bounds are \emph{sorted} from left to right, so we can work with the bounds efficiently. Otherwise,
we need extra time $\O(n \log n)$ in the beginning.

\begin{theorem} \label{thm:brep_int}
The problem $\brep(\int)$ with sorted endpoints of the bounds can be solved in time $\O(n+m)$ where
$n$ is the number of vertices and $m$ is the number of edges.
\end{theorem}

The algorithm of Theorem~\ref{thm:brep_int} is almost the same as the algorithm for $\ext(\int)$
of~\cite{kkv,kkosv}. So the techniques developed for the partial representation extension problem
can be directly applied to more general problems.

\begin{theorem} \label{thm:brep_pint}
The problem $\brep(\pint)$ can be solved in time $\O(n^2)$ where $n$ is the number of vertices.
\end{theorem}

We note that it was already observed in~\cite{kkv} that the classes of proper and unit interval
graphs behave differently with respect to the partial representation problem; unit interval graphs
put additional restrictions is the form of precise rational positions. In~\cite{kkorssv}, the problem
$\ext(\uint)$ was solved in quadratic time by linear programming. So it seemed that this
difference is only in some additional numerical problems posed by unit intervals.
Theorem~\ref{thm:brep_pint} shows together with the result of~\cite[Proposition
2]{kkorssv} that this understanding is fundamentally wrong (unless $\cP = \cNP$, of course):

\begin{theorem}[Klav\'{\i}k et al.~\cite{kkorssv}]
The problem $\brep(\uint)$ is \cNP-complete.
\end{theorem}

The problem is reduced from 3-partition and the hard part is to derive a correct ordering $\blt$ of
the components from left to right; if the ordering is prescribed, one can solve the problem in
quadratic time. The main difference for proper interval graphs is
Proposition~\ref{prop:oncas_blt_ordering} which allows us to derive this ordering $\blt$. The
remainder of the algorithm works similarly as in~\cite{kkorssv}, only some places are more
technical since we have to deal with both left and right bounds; in the case of unit interval
graphs, we can work only with left bounds since the position $\ell(I_v)$ determines the position
$r(I_v)$.

\section{Preliminaries} \label{sec:prelim}

For a graph $G$, we denote by $V(G)$ the set of its vertices and by $E(G)$ the set of its edges. We
use $N[u]$ for the \emph{closed neighborhood} of the vertex $u$, i.e, $N[u] = \{v \in V(G) : uv \in
E(G)\} \cup \{u\}$.

A \emph{(partial) ordering} is a transitive, reflexive and antisymmetric relation. A
\emph{pre-ordering} is just a transitive and reflexive relation, so several elements can be equal in
a pre-ordering. An ordering/pre-ordering is called \emph{linear} if every two elements are
comparable.

For an arbitrary subset $S$ of the real line, we define $\ell(S) = \inf \{x : x \in S\}$ and $r =
\sup \{x : x \in S\}$. By $\lessdot$, we denote the \emph{subset ordering} where $S \lessdot T$ for
two subsets $S$ and $T$ if and only if $r(S) \le \ell(T)$; in other words $S$ is completely on the
left of $T$.

\heading{Endpoint Pre-orderings.}
There are two very natural ways how one can work with intervals and interval representations.  
The first option is to assign to each interval $I$ two rational numbers $\ell(I)$ and $r(I)$. The second option
which we prefer in this paper is just to consider the ordering $<$ of the endpoints as they appear from left
to right. The reason is that this ordering contains all information about intersections of
intervals; precise rational positions are not needed, we can just work with a topology of the
representation. We note that this is not the case of unit interval representations, for which one
has to consider precise rational number positions.

In the case of general interval graphs, one can assume that no two endpoints share their positions.
For bounded representations, this is not true anymore since the bounds might force shared positions.
In this case, $\le$ is a linear pre-ordering, with some sets of endpoints being equal in it.  We say
that an endpoint $z$ is \emph{in between} of $x$ and $y$ if $x \le z \le y$. If two endpoints $x$
and $y$ share position, we denote it by $x=y$, and by $x < y$ we denote that $x$ is strictly on the
left of $y$. It is important to state that if $x < y$, then one can add in between of $x$ and $y$ an
arbitrary number of endpoints in any pre-ordering. If $x=y$, then only endpoints sharing the
position with $x$ and $y$ can be added in between.

If we work with representations just as with left-to-right pre-orderings of the endpoints, then how
can we decide whether the endpoints lie in the bounds? Our assumption on the input is
that we are given a linear pre-ordering of the endpoints of the bounds $\L_v$ and $\R_v$. The solution
gives a bounded representation $\calR$ in the form of a joined pre-ordering $\le$ of the endpoints of the
bounds and the intervals. The bounds constraints just say that $\ell(\L_v) \le \ell(I_v) \le
r(\L_v)$ and $\ell(\R_v) \le r(I_v) \le r(\R_v)$.

\heading{Simplifying Bounds.}
For each interval $I_v$, we want $\ell(I_v) \le r(I_v)$. So we assume each pair $\L_v$ and $\R_v$
satisfies $\ell(\L_v) \le \ell(\R_v)$ and $r(\L_v) \le r(\R_v)$. Otherwise we modify the instance by
putting $\ell(\R_v) := \ell(\L_v)$, resp. $r(\L_v) := r(\R_v)$.

\section{Bounded Representations of Interval Graphs} \label{sec:interval_graphs}

In this section we establish Theorem~\ref{thm:brep_int} which states that the problem $\brep(\int)$
can be solved in time $\O(n+m)$ (given the pre-ordering of the endpoints of the bounds).
First, we give a characterization of bounds for which the bounded representation exists.
Then we describe the algorithm which checks this characterization, and since it is constructive, it
can construct the bounded representation if it exists. We note that our approach is very similar
to~\cite{kkosv}.

\subsection{Characterization of Fulkerson and Gross}

Fulkerson and Gross~\cite{maximal_cliques} gave the following characterization:

\begin{lemma}[Fulkerson and Gross] \label{lem:int_char}
A graph $G$ is an interval graph if and only if there exists a linear ordering $<$ of the maximal
cliques of $G$ such that for every vertex $v \in V(G)$ the cliques containing $v$ appear consecutively
in $<$.
\end{lemma}

\begin{proof}[Sketch]
We sketch this proof since it is important to understand the characterization. Let $\calR$ be an
interval representation.  For each maximal clique $C$, we consider $\bigcap_{v \in C} I_v$, and
according to Helly's theorem this intersection is non-empty. We pick an arbitrary point from this
intersection, and we call it a \emph{clique-point} and denote it by $\cp(C)$. Since these
intersections are for different maximal cliques pairwise distinct, the clique-points are linearly
ordered from left to right. It is routine to check that this is the ordering $<$ from
Lemma~\ref{lem:int_char}.

On the other hand, given an ordering $<$ of the maximal cliques, we place clique-points arbitrarily
in this ordering. Then for each vertex $v$, we put
\begin{equation} \label{eq:cp_construction}
\ell(I_v) = \min \{\cp(C) : v \in C\},
\qquad\text{and}\qquad 
r(I_v) = \max \{\cp(C) : v \in C\},
\end{equation}
i.e., we place $I_v$ on top of the clique-points of cliques containing $v$. We obtain a correct
interval representation of $G$.\qed
\end{proof}

\subsection{Orderings of Maximal Cliques Compatible with the Bounds}

We want to construct a bounded representation in a similar manner, first by placing the
clique-points from left to right and then by constructing the intervals using
(\ref{eq:cp_construction}). But to ensure that the resulting representation is bounded, we cannot
place the clique-points arbitrarily. For a maximal clique $C$, we denote by $J_C$ the set of
possible positions where $\cp(C)$ can be placed; see Appendix~\ref{sec:appendix_int_oncas} for the
precise definition. 

But now if $J_C \lessdot J_C'$, we know that $\cp(C)$ has to be always placed on the left of
$\cp(C')$; so $\lessdot$ on the sets $J_C$ gives the partial ordering of the cliques from
left to right which we denote $\lessdot$ as well.

\begin{proposition} \label{prop:int_oncas}
There exists a bounded representation $\calR$ if and only if there is an ordering $<$ of the maximal
cliques which is consecutive in every vertex $v$ and extends $\lessdot$. 
\end{proposition}

\begin{proof}[Sketch]
The constraints are necessary. We place the clique-points greedily from left to right according to
the ordering $<$. When we place $\cp(C)$, we place it on the right of the previously placed
clique-point and in $J_C$. For contradiction suppose that no such point of $J_C$. We obtain a
contradiction with the consecutivity property or the ordering $\lessdot$. The full proof is in
Appendix~\ref{sec:appendix_int_oncas}.\qed
\end{proof}

\subsection{The Algorithm}

To solve $\brep(\int)$, we proceed in the following main steps.
\begin{packed_enum}
\item We find maximal cliques of $G$, using the algorithm of Rose et
al.~\cite{recog_chordal_graphs} in time $\O(n+m)$.
\item We compute the sets $J_C$, this can be done by a single sweep from left to right in time
$\O(n+m)$. This gives us the partial ordering $\lessdot$ of maximal cliques according to which the
clique-points have to appear on the real line.
\item Test whether there is a linear ordering $<$ of the maximal cliques which extends $\lessdot$
and for each vertex the maximal cliques containing it appear consecutively.  This can be done
using~\cite[Section 2]{kkosv}.
\item If there is a suitable reordering $<$ of $\lessdot$, then we place the clique-points as in the
proof of Proposition~\ref{prop:int_oncas}. Using (\ref{eq:cp_construction}) we construct a correct bounded
representation $\calR$ of $G$.
\end{packed_enum}

Note that if we only want to decide $\brep(\int)$ without constructing a representation, then the
last step can be omitted. For the first step, the input graph has to be chordal and then the total
size of all cliques is $\O(n+m)$.

The constructed representation with $\cp(C) \in J_C$ is correct since we have $\ell(I_v) \in \L_v$
and $r(I_v) \in \R_v$ for each $v \in V(G)$. Moreover $\calR$ is an interval representation of $G$,
since every clique-point lies exactly in the intervals representing the vertices of the
corresponding maximal clique. Thus we can summarize the results.

\begin{proof}[Theorem~\ref{thm:brep_int}]
The proof follows the steps described in the beginning of the section. The correctness of the
algorithm is ensured by Proposition~\ref{prop:int_oncas}. We already observed that for a given
pre-ordering $\le$ of the endpoints of the bounds from left to right, the construction of the
reordering $<$ of $\lessdot$ can be done in time $\O(m+n)$. Since the representation $\calR$ can be
also constructed in linear time with respect to the size of $G$, we see that the whole algorithm
runs in time $\O(m+n)$. \qed
\end{proof}

\section{Bounded Representations of Proper Interval Graphs} \label{sec:pint_graphs}

In this section, we establish Theorem~\ref{thm:brep_pint} which states that the bounded
representation problem of proper interval graphs can be solved in time $\O(n^2)$. 
Proper interval representations give two important orderings: the ordering $\blt$ of the components,
and the ordering $\wlt$ of the intervals of the components. We first describe them in details and then
we show how they can be used in solving of the $\brep(\pint)$ problem.

\subsection{Component Orderings $\blt$}

Let $\calR$ be any representation of $G$ and let $C$ be a connected component. Then $\bigcup_{v \in
C} I_v$ is a closed interval of the real line.  Since the intervals corresponding to the components
are pairwise disjoint, the components are ordered as $C_1 \blt \cdots \blt C_c$. Notice that for
different representations we may get different orderings $\blt$, and when no restriction is posed on
the representation, we can use each of the $c!$ possible orderings.

Suppose that $uv \notin E(G)$. We ask what conditions the bounds have to satisfy to determine that
$I_u \lessdot I_v$ in any bounded representation of $G$. Since the intervals $I_u$ and $I_v$ do not
intersect, it is sufficient to prove that $\ell(I_u)$ is always to the left of $r(I_v)$. This is
clearly satisfied if and only if $\L_u \lessdot \R_v$.

For a given instance of the bounded representation problem, our goal is to determine some ordering
$\blt$ in which a bounded representation exists. To do so, we derive a relation $\blt'$ such
that the ordering $\blt$ of every bounded representation $\calR$ has to extend $\blt'$.
Let $C$ and $C'$ be two distinct components of $G$. We put $C \blt' C'$ if there exists a pair
$u \in C$ and $v \in C'$ such that $\L_u \lessdot \R_v$.

The following proposition (whose proof can be found in
Appendix~\ref{sec:appendix_oncas_blt_ordering}) states that respecting the ordering $\blt'$ is
already sufficient for solving the bounded representation problem:

\begin{proposition} \label{prop:oncas_blt_ordering}
A bounded representation of $G$ in an ordering $\blt$ extending $\blt'$ exists if and only if there
exists a bounded representation of $G$.
\end{proposition}

\begin{proof}[Sketch]
We argue only the non-obvious direction. 
Suppose that $C$ and $C'$ are two
components incomparable in $\blt'$. In such a case, their bounds have to be hugely overlapping.
There are two cases one has to deal with:
\begin{packed_itemize}
\item All bounds of $C$ and $C'$ are pairwise intersecting. Then due to Helly's theorem, we can represent $C$ and $C'$ in
any ordering in this intersection.
\item Only bounds of, say, $C$ are pairwise intersecting. But then due to Helly's theorem, we can
represent $C$ either on the left of the left-most bound of $C'$, or on the right of the right-most
bound of $C$. We still leave enough space for $C'$ to be represented.
\end{packed_itemize}
Then we repeatedly apply this local reordering of incomparable components till we modify the given
bounded representation in the prescribed ordering $\blt$ which extends $\blt'$. See
Appendix~\ref{sec:appendix_oncas_blt_ordering} for further details.\qed
\end{proof}

We note that a similar proposition is not correct for unit interval graphs. The problem is that a
component has some minimal size which it requires in every representation, so it cannot be placed in
this arbitrary small common intersection of the bounds. Actually Klav\'{i}k et al.~\cite[Theorem
1]{kkorssv} proved that finding the correct ordering $\blt$ is the \cNP-complete part of the problem
$\brep(\uint)$. For a prescribed ordering $\blt$, one can solve the bounded representation problem
of unit interval graphs in quadratic time.

\subsection{Vertex Orderings $\wle$}

Two vertices $u$ and $v$ are called \emph{indistinguishable} if $N[u] = N[v]$. So being
indistinguishable defines an equivalence relation on $V(G)$, and the classes of this equivalence are
called \emph{groups of indistinguishable vertices}. For every intersection representation, the
vertices of each group can be represented the same, and so indistinguishable vertices are not very
interesting from the structural point of view. This is not the case for the bounded representation
problem (or any other problem of restricted representation), in which indistinguishable vertices can
be given distinct bounds and thus are forced to be represented differently.

\heading{Vertex Orderings.}
Let $\calR$ be any proper interval representation, and assume for a second that no two intervals of
$\calR$ are the same. Then the intervals are ordered from left to right, and we denote this ordering by
$\wlt$. The ordering $\wlt$ is the ordering of the left endpoints, and at the same time the ordering
of the right endpoints. In $\wlt$, each group of indistinguishable vertices has to appear
consecutively. Deng et al.~\cite{deng} characterize possible orderings $\wlt$ for connected proper
interval graphs:

\begin{lemma}[Deng et al.]
For a connected proper interval graph, the ordering $\wlt$ is uniquely determined up to local
reordering of the groups of indistinguishable vertices and the complete reversal.
\end{lemma}

In other words, there exists a partial ordering $<$ in which exactly the pairs of indistinguishable
vertices are incomparable. Then each $\wlt$ is a linear extension of $<$ or its reversal. Corneil et
al.~\cite{uint_corneil} describe a simple linear-time algorithm for computing $<$.

Now we allow having several same intervals in the representation $\calR$ since the bounds might
force this situation. The representation $\calR$ then gives a linear pre-ordering $\wle$. When we construct
bounded representations, we place intervals as the same if and only if this is forced by the bounds.
It is easy to observe that if $I_u = I_v$, then the vertices $u$ and $v$ are indistinguishable. 

\heading{Constraints Given by Bounds.}
In the case of bounded representations, the order of some pairs of the indistinguishable vertices
can be prescribed by the bounds. Suppose that we restrict ourself to just a single component $C$ of
the input graph $G$ and ignore the rest.  Similarly to above, we produce a relation $\wlt'$
of the vertices of $C$.

Let $u$ and $v$ be two indistinguishable vertices of $C$. We put $u \wle' v$ if and only if $\L_u
\lessdot \L_v$ or $\R_u \lessdot \R_v$; so $\wle'$ is a union of the subset order $\lessdot_\ell$ of
the left bounds and the subset order $\lessdot_r$ of the right bounds. Notice that the pre-ordering
$\wle'$ does not have to be a partial ordering and that $u \wle' v$ implies $u \wle v$ for any
representation $\calR$.

Now, since we do not want to work with pre-ordering $\wle$, we construct a reduced graph $C'$ with
modified bounds. The following proposition states that this construction does not change solution of
the problem.

\begin{proposition} \label{prop:oncas_wlt_ordering}
There exists a bounded representation of $C$ with an ordering extending $<$ if and only if
there exists a bounded representation of $C'$ in an ordering $\wlt$ which
extends both $<$ and $\wlt'$.
\end{proposition}

\begin{proof}[Sketch]
The construction of $C'$ is done in two steps. First, we consider strongly connected components
defined by $\wle'$, and they have to be represented by the same intervals. Therefore, we unify the
bounds of their intervals to force this.

To prove the correctness of the reduction, we reorder groups in $C$ according to $\wlt$. For each
group, we apply a similar greedy procedure as in Proposition~\ref{prop:int_oncas}. For the detailed
proof see Appendix~\ref{sec:appendix_oncas_wlt_ordering}.\qed
\end{proof}

\subsection{The Algorithm}

The algorithm works as follows:

\begin{packed_enum}
\item We compute the ordering $\blt'$ of components, and construct a linear ordering $\blt$
extending $\blt'$.
\item We proceed the components according to $\blt$ from left to right: $C_1 \blt \cdots \blt C_c$.
\item When processing the component $C_i$:
\begin{packed_itemize}
\item Compute the partial ordering $<$, using~\cite{uint_corneil}.
\item For $<$ and its reversal do the following: for each group $\Gamma$ of indistinguishable
vertices, compute $\wlt'$, its strongly connected components, the reduced graph $C'_i$ and its
ordering $\wlt$.
\item Place the endpoints according to $\wlt$ from left to right, on the right side of the
representation of $C_{i-1}$ greedily as far to the left as possible.
\item Construct a representation of $C_i$, by copying the intervals $I_{S_i}$.
\end{packed_itemize}
\end{packed_enum}

It remains to argue details concerning specific implementation and correctness which is easily
implied by Proposition~\ref{prop:oncas_blt_ordering} and Proposition~\ref{prop:oncas_wlt_ordering}.
See Appendix~\ref{sec:appendix_brep_pint} for details.

\section{Conclusions} \label{sec:conclusions}

In this paper, we give a polynomial time algorithm for the classes of interval and proper interval
graphs for a recently introduced problem $\brep$. The main result of this paper is a rather
surprising discovery that the bounded representation problem distinguishes the classes of proper and
unit interval graphs: $\brep(\pint)$ is polynomially solvable but $\brep(\uint)$ is
\cNP-complete~\cite{kkorssv}. We believe that is a very interesting problem to further investigate
differences between the structures of proper and unit interval representations; this paper gives a
good reason to do so.

\heading{Open Problems.} We conclude with two open problems.

\begin{problem}
Is it possible to solve $\brep(\pint)$ in time $\O(n+m)$ (with a given left-to-right ordering of the
bounds)?
\end{problem}

The current bottleneck of our algorithm is the computation of $\wle$ from $\wle'$ which is the only
step requiring time $\O(n^2)$.

\begin{problem}
What is the complexity of the $\brep$ problem for other classes such as circular-arc graphs, circle
graphs?
\end{problem}

Currently, the only known results are for the classes $\int$, $\pint$, and $\uint$. Even attacking
some simpler problems for these classes might be very interesting. For instance, solving the partial
representation extension problem for circular-arc graphs could be a major advancement in the area of
the restricted representation problems.

\bibliographystyle{splncs03}
\bibliography{bounded_int_rep}

\appendix

\section{Appendix: Proof of Proposition~\ref{prop:int_oncas}} \label{sec:appendix_int_oncas}

Let $G$ be a given graph. For a maximal clique $C \subseteq G$, we set
$$J_C=\left( \bigcap_{u \in V(C)}[\ell(\L_u),r(\R_u)]
\right) \setminus \left( \bigcup_{v \notin V(C)}[r(\L_v),\ell(\R_v)]\right)$$i
to denote the set of all the possible candidates for clique points $\cp(C)$ with respect to the
given restrictions on the intervals of the representation.  (Where the interval
$[r(\L_v),\ell(\R_v)]$ is empty if $r(\L_v) > \ell(\R_v)$.) If the set $J_C$ is empty, then it is
clearly not possible to place $\cp(C)$ and thus there is no interval representation of $G$ which
satisfies the given restrictions.

We now prove Proposition~\ref{prop:int_oncas} which says that there exists a bounded representation
$\calR$ if and only if there is an ordering $<$ of maximal cliques which is consecutive in every
vertex $v$ and extends the subset ordering $\lessdot$ on the sets $J_C$ (which can be also
understood as a partial ordering of the maximal cliques).

\begin{proof}[Proposition~\ref{prop:int_oncas}]
If there is no such extension $<$, then using Lemma~\ref{lem:int_char}
we see that $G$ does not have an interval representation which satisfies the bounds, 
because the conditions forced by $\lessdot$ are clearly necessary.

On the other hand, suppose that $G$ is a graph which has an interval representation satisfying the bounds. Then according to
Lemma~\ref{lem:int_char} there is a reordering $<$ which extends $\lessdot$.
To construct an interval representation $\calR$ of $G$ we
first greedily place the clique points according to $<$ from left to right always as far to the left
as possible.

All that is left is to show that the greedy procedure cannot fail. Assume, for the sake of
contradiction, that the procedure fails for the clique point $\cp(C)$. Since $\cp(C)$ cannot be
placed at all, there are some clique points placed on the right of $r(J_C)$ (or possibly on
$r(J_C)$). Let $\cp(B)$ be the leftmost of them. Since $\cp(B)$ was placed before $\cp(C)$ we have
$B<C$ and thus $C\lessdot B$ cannot hold. Therefore we know that $ \ell(J_{B})<r(J_C)$.

Following the greedy procedure we see that $\cp(B)$ was not placed to the left of $r(J_C)$, because
all the possible locations there were blocked by previously placed clique points or by intervals
$[r(\L_v),\ell(\R_v)]$ where $v \notin V(B)$. There is at least one clique point placed to the right of
$\ell(J_{B})$, because otherwise we would place $\cp(B)$ to $\ell(J_B)$ or right next to it. Let $\cp(A)$
be the rightmost clique-point placed between $\ell(J_{B})$ and $\cp(B)$.

We claim that every point between $\cp(A)$ and $r(J_C)$ has to be covered by intervals
$[r(\L_v),\ell(\R_v)]$ where $v \notin V(B)$. Otherwise we would place $\cp(B)$ to the uncovered point,
which is in $J_B$, since $\cp(A)$ is between $\ell(J_B)$ and $\cp(B)$. Let $S$ be the set of such
intervals and let $\calC$ be the set of maximal cliques containing at least one vertex from
$S$. Note that since $S$ induces a connected subgraph of $G$, all the cliques in $\calC$
appear consecutively in $<$, because every pair of adjacent vertices from $S$ is contained in some
maximal clique from $\calC$.

From the assumptions made, it is clear that $A$ and $C$ are in $\calC$, but $B$ is not and
$A<B<C$ holds. However consecutiveness of $\calC$ and $A<B$ imply $C<B$ which gives us a
contradiction.\qed
\end{proof}

All that is left is to describe the construction of a suitable interval representation of $G$ provided that the partial ordering $\lessdot$ was extended to $<$ in the third step of the algorithm.

To obtain such representation $\calR$ it suffices to let \[\ell(I_v)=\min\left(\{r(\L_v)\} \cup \{\cp(C) \mid v \in V(C)\}\right)\] and  \[r(I_v)=\max\left(\{\ell(\R_v)\} \cup \{\cp(C) \mid v \in V(C)\}\right)\] for every vertex $v$ of $G$. 

Then from the fact that $\cp(C) \in J_C$ we have $\ell(I_v) \in \L_v$ and $r(I_v) \in \R_v$. Therefore the restrictions on the representing intervals are satisfied. Moreover $\calR$ is an interval representation of $G$, since every clique-point lies exactly in the intervals representing the vertices of the corresponding maximal clique.

\section{Appendix: Proof of Proposition~\ref{prop:oncas_blt_ordering}} \label{sec:appendix_oncas_blt_ordering}

In this part we prove that a  bounded representation of $G$ in an ordering $\blt$ extending $\blt'$ exists if and only if there
exists a bounded representation of $G$.

An interval $I$ is called \emph{trivial} if $\ell(I) = r(I)$ and \emph{non-trivial} otherwise.
Two non-trivial intervals $I$ and $J$ are \emph{intersecting non-trivially}, if $I \cap J$ is
non-trivial. If $I$ is trivial and $J$ is non-trivial, then $I \cap J$ is \emph{non-trivial} if
$\ell(J) < \ell(I) = r(I) < r(J)$.

Before proving Proposition~~\ref{prop:oncas_blt_ordering}, we need to establish some basic properties.  For the following, let
$C$ and $C'$ be two incomparable components in $\blt'$. Let $\calB$ be the collection of all bounds
of intervals in $C$, and similarly $\calB'$ for $C'$.

\begin{lemma} \label{lem:incomparable}
Every bound $B \in \calB$ non-trivially intersects every bound $B' \in \calB'$.
\end{lemma}

\begin{proof}
If $B$ is $\L_u$ and $B'$ is $\R_v$, or vice versa, then the
statement clearly holds; if $B$ and $B'$ would not intersect, or would intersect trivially, we get
that $\L_u \lessdot \R_v$ and $C \blt' C'$. Now, say we have two left bounds $B = \L_u$ and $B' = \L_v$, and
suppose for contradiction $\L_u \lessdot \L_v$.  Since $\ell(\L_v) \le \ell(\R_v)$, we get that also
$\L_u \lessdot \R_v$ and so $C \blt' C'$, contradiction. We proceed similarly for the two right
bounds.\qed
\end{proof}

\begin{lemma} \label{lem:nonint_C}
If the bounds $\calB$ are not pairwise non-trivially intersecting, then the bounds $\calB'$ are pairwise
non-trivially intersecting.
\end{lemma}

\begin{proof}
Let $\L_u$ be the leftmost left bound of $C$ (minimizing $r(\L_u)$) and let $\R_v$ be the rightmost
right bound of $C$ (maximizing $\ell(\R_v)$). We know that $\L_u \lessdot \R_v$. If $B' \in
\calB'$, then according to Lemma~\ref{lem:incomparable} we have $\ell(B') < r(\L_u) \le \ell(\R_v)
< r(B')$. So all bounds of $\calB'$ are intersecting non-trivially.\qed
\end{proof}

\begin{lemma} \label{lem:swapping}
Suppose that there exists a bounded representation $\calR$ which places $C$ and $C'$ next to each
other in $\blt$. Then there exists another bounded representation $\calR'$ with the only difference
that the order of $C$ and $C'$ is swapped.
\end{lemma}

\begin{proof}
We represent the remaining components in $\calR'$ exactly as in $\calR$, so we only need to deal
with $C$ and $C'$. We can assume that $C \blt C'$ in $\calR$. Let $x$ be the rightmost endpoint of
the component on the left of $C$, and let $y$ be the leftmost endpoint of the component on the right
of $C$. (Or $-\infty$, resp. $\infty$, if such an endpoint does not exist.)

There are three possible cases how the bounds $\calB$ and $\calB'$ are intersecting.
\begin{packed_itemize}
\item {\bf Case 1:} The bounds $\calB$ are pairwise non-trivially intersecting and the bounds
$\calB'$ are as well. Then by Lemma~\ref{lem:incomparable} all bounds of
$\calB \cup \calB'$ are pairwise non-trivially intersecting. So by the Helly property, there exists a
non-trivial interval $J$ contained in all bounds $\calB \cup \calB'$ such that $x < \ell(J) < r(J) <
y$. Since $\calR$ exists, there exists some ordering of the endpoints in which the components $C$
and $C'$ are representable. We can represent in $\calR'$ the components $C$ and $C'$ in this
ordering inside $J$ which gives a correct bounded representation of $G$.
\item {\bf Case 2:} The bounds $\calB$ are not pairwise non-trivially intersecting and the bounds
$\calB'$ are. Let $\L_u$ and $\R_v$ be the same as in the proof of Lemma~\ref{lem:nonint_C}. Also,
in the proof we argued that there exists a non-trivial interval $J$ contained in every bound of
$\calB'$ such that $\ell(J) < r(\L_u) \le \ell(\R_v) < r(J)$. So it is possible to place the entire
representation of $C'$ strictly in between of $x$ and $r(\L_u)$, in the same ordering as in $\calR$.
And we represent $C$ on the right of it, again in the same ordering. (This is possible since each
bound ends on the right of $r(\L_u)$.)
\item {\bf Case 3:} The same as above, but we represent $C$ on the right of $J$, and we compress
$C'$ on the left of it.
\end{packed_itemize}

For the cases 1 to 3, we construct a correct bounded representation $\calR'$ in the required
ordering $\blt$. Lemma~\ref{lem:nonint_C} states that Case 4~in which both $\calB$ and $\calB'$ are
not pairwise non-trivially intersecting cannot occur.\qed
\end{proof}

\begin{proof}[Proposition~\ref{prop:oncas_blt_ordering}]
The implication from left to right is obvious, $\calR$ is one of the bounded representations of
$G$. For the other implication, let $\overline\blt$ be the ordering of the components in
$\overline\calR$, and modify $\overline\calR$ by repeated application of Lemma~\ref{lem:swapping} to
make it in the ordering $C_1 \blt \cdots \blt C_c$. Now suppose that $\overline\calR$ starts by
$C_1,\dots,C_{i-1}$ but the following is not $C_i$ as in $\blt$. We take $C_i$ in $\overline\blt$
and by repeated application of Lemma~\ref{lem:swapping}, we shift it next to $C_{i-1}$. To do so, we
need to show that $C_i$ is incomparable to all components between $C_{i-1}$ and $C_i$ in
$\overline\blt$. But if there would be $C_j$ which would be comparable, then $C_j \blt' C_i$, and so
the ordering $\blt$ would not extend $\blt'$.\qed
\end{proof}

\section{Appendix: Proof of Proposition~\ref{prop:oncas_wlt_ordering}} \label{sec:appendix_oncas_wlt_ordering}

Let us remind the statement of Proposition~\ref{prop:oncas_wlt_ordering}. It says that there exists a bounded representation of $C$ with an ordering extending $<$ if and only if there exists a bounded representation of $C'$ in an ordering $\wlt$ which extends both $<$ and $\wlt'$. Before the proof of this statement we need to introduce some notation.

\heading{Reduced Graph.}
We have a pre-ordering $\wle'$, and some intervals might be forced to be represented the same. We
are going to construct a reduced graph $C'$ such that it can be represented in a strict ordering
$\wlt$. We consider one group $\Gamma$ of indistinguishable vertices and use the relation $\wlt'$ to
define an oriented graph $H$ on $\Gamma$; we put $(u,v) \in E(H)$ if and only if $u \wlt' v$. Let
$S$ be a strongly connected component of $H$. Then its vertices have to be represented by the same
interval in $\calR$. The strongly connected components are partially ordered by the remaining edges
of $H$ going in between of them, let $S_1 < \dots < S_k$ be an arbitrary topological sorting. Then
the constructed pre-ordering $\wlt$ orders the vertices of $\Gamma$:
$$S_1 \wlt S_2 \wlt \cdots \wlt S_k,$$
where all vertices of each $S_i$ are equal in $\wle$.

Let $C'$ be the contracted graph, in which vertices of each group are replaced by the vertices
$S_1,\dots,S_k$. The idea is to force the vertices of each $S_i$ to be represented by the same
interval. We are going to represent the entire $S_i$ by a single interval $I_{S_i}$. To do that we need to
define bounds compatible with all vertices of $S_i$, and so we put 
$$\L_{S_i} = \bigcap_{u \in S_i} \L_u,\qquad\text{and}\qquad\R_{S_i} = \bigcap_{u \in S_i} \R_u.$$
Then the constructed pre-ordering $\wle$ becomes a linear ordering $\wlt$ for $C'$, so the representation
of $C'$ has pairwise distinct intervals.  So in the constructed representation of $C'$, we can
replace the interval $I_{S_i}$ with several equal intervals, representing the vertices of $S_i$, and
we obtain a correct bounded representation of $C$.

\heading{Group Reordering.}
Before proving this Proposition~\ref{prop:oncas_wlt_ordering}, we state one important property. Let $\Gamma$ be a group of
indistinguishable vertices and let $\calR$ be a representation with the ordering $\wlt$. Notice
that the vertices of $\Gamma$ appear consecutively in $\wlt$. We study which points of the real line
are taken by all intervals of $\Gamma$ and which by some interval of $\Gamma$. We define
$$
\medcap\Gamma = \bigcap_{v \in \Gamma} I_v,
\qquad\text{and}\qquad
\medcup \Gamma = \bigcup_{v \in \Gamma} I_v.
$$
Since intervals of the real line satisfy the Helly property and $\Gamma$ forms a clique in $G$, we get
that both $\medcap \Gamma$ and $\medcup \Gamma$ are closed intervals such that $\medcap\Gamma
\subseteq \medcup \Gamma$.

\begin{lemma}
If $u$ is adjacent to $\Gamma$, then $I_u$ intersects $\medcap \Gamma$. And if $u$ is non-adjacent
to $\Gamma$, then $I_u \cap \medcup \Gamma = \emptyset$.
\end{lemma}

\begin{proof}
Since $\Gamma$ is a group of indistinguishable vertices, then $u$ is either adjacent to everything
or nothing, the rest easily follows.\qed
\end{proof}

Let $\Gamma$ and $\Gamma'$ be two groups. If the vertices in these groups are pairwise adjacent, we
get that $\medcap \Gamma$ intersects $\medcap \Gamma'$, and if they are pairwise non-adjacent, we
get that $\medcup \Gamma \cap \medcup \Gamma'$ is empty.

\begin{lemma} \label{lem:group_reordering}
Let $\calR$ be a representation of $G$ and let $\Gamma$ be a group of indistinguishable vertices. Let $\calR'$ be a
representation constructed from $\calR$ by placing intervals $I'_u$ of $\Gamma$ such that
$\medcap \Gamma \subseteq I'_u \subseteq \medcup \Gamma$ with the same ordering of the left and
right endpoints. Then $\calR'$ is a correct proper interval representation of $G$.
\end{lemma}

\begin{proof}
The representation $\calR'$ is proper, and it is easy to see that intersections are preserved.\qed
\end{proof}

\begin{proof}[Proposition~\ref{prop:oncas_wlt_ordering}]
Again, one direction is clear. We can construct representation of $C$ as above by copying the
intervals $I_{S_i}$. For the other direction, let $\wlt$ be an ordering from the
statement and let $\calR$ be a representation of $C$. We proceed the groups of $C$ in an arbitrary order
and construct a representation of $C'$ according to $\wlt$ using Lemma~\ref{lem:group_reordering}.

For a group $\Gamma$ with the vertices $S_1 \wlt \cdots \wlt S_k$, we want to place the
intervals $I_{S_i}$ in between of $\medcap \Gamma$ and $\medcup \Gamma$ in this order. Let $\ell_i$
(resp. $r_i$) be the left (resp. right) endpoint of $I_{S_i}$.  We want to place the left endpoints
in $[\ell(\medcup \Gamma),\ell(\medcap \Gamma)]$ and the right endpoint in $[r(\medcap
\Gamma),r(\medcup \Gamma)]$ both according to the ordering $\wlt$, and we can do this independently.

We consider the following only for the left endpoints $\ell_1,\dots,\ell_k$ and we place the right
endpoints $r_1,\dots,r_k$ in $[r(\medcap \Gamma),r(\medcup \Gamma)]$ using exacly the same argument.
Let $\L_1,\dots,\L_k$ be the left bounds of $S_1,\dots,S_k$. We assume that each $\L_i$ is a subset
of $[\ell(\medcup \Gamma),\ell(\medcap \Gamma)]$; otherwise we consider only the intersections of
the bounds with this interval. (Notice that even for these restricted bounds we get the constraints
$\wlt'$.) We place the left endpoints in $[\ell(\medcup \Gamma),\ell(\medcap \Gamma)]$ greedily from
left to right from $\ell_1$ to $\ell_k$ while respecting their bounds.

Suppose that this placing procedure fails when we attempt to place $\ell_i$. Then we show that there is a
contradiction with $\wlt$ extending $\wlt'$. Since we successfully placed $\ell_{i-1}$ and we cannot
place $\ell_i$, we have $r(\L_i) \le \ell_{i-1}$.  The reason why $\ell_{i-1}$ was not placed more
to the left by the greedy algorithm is that this position is blocked by some previously placed left
endpoint. Let $\ell_j$ be the leftmost placed endpoint such that $r(\L_i) \le \ell_j$. Since
$\ell_j$ was placed as far to the left as possible and the position immediately on the left of
$r(\L_i)$ is not blocked by any other left endpoint, we get $r(\L_i) \le \ell(\L_j)$.

It remains to relate this to the original bounds of $C$, and to show that it contradicts the definition
of $\wle$. Let $u \in S_i$ be a vertex such that $r(\L_u)$ is minimal and choose $v \in S_j$ such that
$\ell(\L_v)$ is maximal. Then we get $r(\L_u) \le \ell(\L_v)$, and so $u \wlt' v$. This contradicts
the constructed topological sort which places $S_j$ before $S_i$, since there is an edge going from
$S_i$ to $S_j$.\qed
\end{proof}

\section{Appendix: Proof of Theorem~\ref{thm:brep_pint}} \label{sec:appendix_brep_pint}

In this part the reader can find details of the proof of Theorem~\ref{thm:brep_pint}, which says that the problem $\brep(\pint)$ can be solved in time $\O(n^2)$ where $n$ is the number of vertices.

\begin{lemma} \label{lem:computing_blt}
A linear ordering $\blt$ extending $\blt'$ can be computed in time $\O(n+m)$.
\end{lemma}

\begin{proof}
Let $C_1,\dots,C_c$ be the components of $G$. We define for each component two numbers called
a \emph{lower handle} and an \emph{upper handle}:
$$\LH(C_i) = \min\{r(\L_v) : v \in V(C_i)\},
\quad\text{and}\quad
\UH(C_i) = \max\{\ell(\R_v) : v \in V(C_i)\}.$$
In this setting, we get that $C_i \blt' C_j$ if and only if $\LH(C_i) \le \UH(C_j)$.

It is proved in~\cite[Section 2.2]{kkosv} that one can find a linear ordering $\blt$ extending
$\blt'$ using these handles. We first compute a linear ordering $\prec$ of all the handles from left
to right, and to deal with ties we first place the lower handles in any order and then the upper
handles in any order. Since the endpoints of the bounds are given sorted, we can compute this using
a single sweep from left to right.

Now we use this ordering and construct $\blt$ by repeated finding of minimal elements. We look at
the first element in $\prec$. If it is a lower handle $\LH(C_i)$, then there cannot be any other lower
handle $\LH(C_j)$ such that $\LH(C_j) \prec \UH(C_i)$, and then $C_i$ is a minimal element. And if
the first element is an upper handle $\UH(C_i)$, then $C_i$ is a minimal element. If there is no
minimal element, the algorithm fails. If there is some minimal element $C_i$, we append it to the
constructed $\blt$ and remove both handles of $C_i$ from $\prec$. In total, this algorithm can be
implemented in time $\O(n+m)$. For details and a proof of correctness, see~\cite{kkosv}.\qed
\end{proof}

\begin{lemma} \label{lem:computing_wlt}
For a partial ordering $<$ (resp. its reversal) of~\cite{uint_corneil}, we can compute any $\wlt$
extending $<$ (resp. its reversal) and $\wlt'$ in time $\O(n^2)$.
\end{lemma}

\begin{proof}
We argue only for $<$, for reversal the argument is the same. We proceed separately for each group
$\Gamma$, for which $<$ gives no restriction. We represent the constraints given by $\wlt'$ as
an oriented graph $H$ over $\Gamma$, as described above. We start by finding strongly connected
components $S_1,\dots,S_k$. Then we contract the strongly connected components, and find any
topological sorting of the contracted graph. Everything can be done in linear time with respect to the size of the
graph $H$ which is $\O(n^2)$.\qed
\end{proof}

Consider all representations of the reduced graph $C'_i$ in an ordering $\wlt$.  We call a bounded
representation $\calR$ of $C'_i$ in the ordering $\wlt$ the \emph{left-most representation}, if it
minimizes the right-most endpoint of $C'_i$ over all bounded representations $\calR'$ of $C'_i$. To
be more precise, we denote the right-most endpoint of the component $C'_i$ by $r(C'_i)$ in $\calR$
and by $r'(C'_i)$ in $\calR'$. Then $\calR$ is the left-most representation if $r(C'_i) \le
r'(C'_i)$ for every bounded representation of $\calR'$.  Since we are working with left-to-right
orderings instead of precise rational positions, we just require that for any bound $x$ such that
$r'(C'_i) \le x$, we also have $r(C'_i) \le x$.

\begin{lemma} \label{lem:leftmost_rep}
If there exists a representation of the reduced graph $C'_i$ in an ordering $\wlt$, then the left-most representation
exists, it is unique and it can be computed in time $\O(n)$.
\end{lemma}

\begin{proof}
For a given ordering $\wlt$, there are only finitely many ways how one can insert the bounds into
the left-to-right ordering of the endpoints of the bounds; so there are finitely many different
bounded representations. We consider the structure $\rset$ of all bounded representations in the ordering
$\wlt$, and we know that there is at least one such representation.  We are going to show that they
form a meet semilattice with infimum operation defined as follows. Then the left-most representation
corresponds to the infimum of the entire semilattice, and thus we know it always exists and it is
unique.

For two representations $\calR, \calR' \in \rset$, we define the infimum $\calR \wedge \calR'$ as the
representation $\overline\calR$ such that $\overline\ell_i = \min\{\ell_i,\ell'_i\}$ and
$\overline r_i = \min\{r_i,r'_i\}$. In other words, the lattice is defined by the ordering $\le$ in which
$\calR \le \calR'$ if and only if it is less or equal in each endpoint, and then this is the natural
way to define infimum.

It is quite clear that $\overline\calR$ satisfies the bounds. Also, it is straightforward to check
that $\overline\calR$ is a correct proper interval representation of $C'_i$; for each vertex we have
two possible orderings $\ell \ell r r$ or $\ell r \ell r$ if we place $\calR$ together with
$\calR'$. So for a pair of vertices $u$ and $v$, we have four possibilities in total and we just need
to check that the intersections are preserved for all of them.

We can construct the left-most representation in time $\O(n)$ as follows. First, we compute a common
ordering of the left and right endpoints from left to right. We know from $\wle$ that $\ell_1 \le
\cdots \le \ell_n$. Into these orderings, we insert right endpoints one-by-one, and we insert $r_i$
right before the left endpoint $\ell_j$ where $v_j$ is the smallest non-neighbor of $v_i$ such that
$v_i \wlt v_j$. This ordering is uniquely determined by $\wle$ and any representation constructed
in this ordering is a correct representation of $G$.

Now, we process the endpoints from left to right and we always place them  greedily as far to the left
as possible. Doing so, we respect the bounds, so $\ell_i$ is inserted only in $\L_{v_i}$ and $r_i$ only
in $\R_{v_i}$. If we insert a right endpoint $r_i$, and the previously inserted endpoint was
$\ell_j$, then we put $r_i = \min\{\ell_j,\ell(\R_i)\}$. Otherwise, we insert the endpoint $x$
strictly on the right of the previously inserted endpoint $y$ such that $x \ge \ell(B)$ where $B$ is the
bound for $x$; so we might insert it immediately next to $y$, or at the position $\ell(B)$ if the
bound is further to the right.

Since the representation is constructed according to the common ordering, it is a correct proper
interval representation of $G$. We prove by induction that the representation is the left-most
representation, and thus it satisfies the right endpoints of the bounds. (The left endpoints are
clearly satisfied by the construction.) The first endpoint is placed directly on the bound, so it
is clearly the left-most. Now suppose that after placing an endpoint $x$ the representation is not
the left-most anymore, but it was the left-most right before placing the endpoint $x$. But this is not
possible since we clearly place $x$ as far to the left as possible, while satisfying the common
ordering and the left endpoints of the bounds.\qed
\end{proof}

\begin{proof}[Theorem~\ref{thm:brep_pint}]
We are ready to prove the main theorem. Clearly, if no bounded representation exists, the algorithm
has to fail in some step and does not construct it. On the other hand, suppose that some convenient
representation exists. First, we compute the ordering $\blt$ extending $\blt'$
using Lemma~\ref{lem:computing_blt}. According to Proposition~\ref{prop:oncas_blt_ordering}, if
there exists a bounded representation of $G$, then there exists a bounded representation in this
ordering $\blt$. If $\calR$ is a such representation, then the algorithm constructs another bounded
representation $\calR'$.

 We are going to show by induction that $r'(C_i) \le r(C_i)$ for each component $C_i$.
(Since we are working in the context of left-to-right orderings, this condition just means that for
any endpoint $x$ of any bound, if $r(C_i) \le x$, then also $r'(C_i) \le x$.) This condition says
that the representation $\calR'$ of $C_i$ leaves at least as much space as $C_i$ in $\calR$ for the
remaining components.

The first induction step is clearly satisfied for non-existent component $C_0$. (This means that we
can place $C_1$ in both $\calR'$ and $\calR$ arbitrarily to the left.) We proceed the components
from left to right. Let $C_i$ be the processed component, so we have represented $C_1,\dots,C_{i-1}$
in $\calR'$, and the induction hypothesis states that $r'(C_{i-1}) \le r(C_i)$.  We test for $C_i$
both $<$ and its reversal and choose the representation which minimizes $r'(C_i)$.

The component $C_i$ is represented in $\calR$ in an ordering extending one of these orderings, without loss of
generality $<$. Using Lemma~\ref{lem:computing_wlt}, we can compute in time $\O(n^2)$ the reduced
graph $C_i'$ and some other ordering $\wlt$. According to Proposition~\ref{prop:oncas_wlt_ordering},
there exists a representation $\overline\calR$ of $C'_i$, and by copying the intervals $I_{S_i}$, we obtain a
representation $\overline\calR$ of $C_i$. Since we represent each group $\Gamma$ in between of $\medcap
\Gamma$ and $\medcup \Gamma$, the representation $\overline\calR$ of $C_i$ is at most as large as
the representation $\calR$, and thus $\overline r(C_i) \le r(C_i)$.

The proof of Proposition~\ref{prop:oncas_wlt_ordering} is just existential, since it requires a
representation $\calR$. However using Lemma~\ref{lem:leftmost_rep}, we can construct the left-most
representation $\calR'$ of $C'_i$ which satisfies $r'(C'_i) \le \overline r(C'_i) = \overline
r(C_i)$. So by copying the intervals $I_{S_i}$, we obtain a representation $\calR'$ of $C_i$ which
satisfies $r'(C_i) = r'(C'_i) \le r(C_i)$ as required. Since at least for $<$ or its reversal we
obtain a representation $\calR'$ satisfying $r'(C_i) \le r(C_i)$, the induction step is correct
and we construct a correct bounded representation of $G$ if it exists.\qed
\end{proof}

\end{document}